\documentclass[preprint,showpacs,preprintnumbers,amsmath,amssymb]{revtex4}
\usepackage{graphicx}
\usepackage{bm}
\usepackage{amsmath}
\begin{document}

\title{Rotational Symmetry Breaking in Sodium Doped Cuprates}

\author{Yan Chen$^1$, T. M. Rice$^{1,2}$, and F. C. Zhang$^{1}$}

\affiliation{$^1$Department of Physics and Center of Theoretical and
Computational Physics, The University of Hong Kong, Pokfulam Road,
Hong Kong, China\\ $^2$Institut f\"ur Theoretische Physik,
ETH-Z\"urich, CH-8093 Switzerland}

\date{\today}

\begin{abstract}
For reasonable parameters a hole bound to a Na$^{+}$ acceptor in
Ca$_{2-x}$Na$_{x}$CuO$_{2}$Cl$_{2}$ has a doubly degenerate ground
state whose components can be represented as states with even (odd)
reflection symmetry around the $x$($y$) -axes. The conductance
pattern for one state is anisotropic as the tip of a tunneling
microscope scans above the Cu-O-Cu bonds along the $x$($y$)-axes.
This anisotropy is pronounced at lower voltages but is reduced at
higher voltages. Qualitative agreement with recent experiments leads
us to propose this effect as an explanation of the broken local
rotational symmetry.

\end{abstract}

\pacs{74.72.-h, 74.62.Dh, 74.25.Jb}

 \maketitle

The recent atomically resolved scanning tunneling microscopy (STM)
studies by Davis and collaborators~\cite{STM1,STM2} on strongly
underdoped Ca$_{2-x}$Na$_{x}$CuO$_{2}$Cl$_{2}$ revealed a
surprisingly complex pattern with the square symmetry of the lattice
broken on a local scale. The STM data were analyzed to obtain the
local hole density of Cu-site using the method proposed by Randeria,
Sensarma, Trivedi and Zhang~\cite{Ratio1} and also a related method
proposed by Anderson~\cite{Ratio2}. In these methods, the
differential conductance signal is integrated from the chemical
potential to a substantial voltage cutoff. Randeria {\em et al.}
~\cite{Ratio1} pointed out that the ratio between its positive
voltage (electron injection) and negative voltage (hole injection)
signals measured the local hole density of the Cu-site independent
of the strength of the troublesome tunneling matrix elements.

In Ca$_{2-x}$Na$_{x}$CuO$_{2}$Cl$_{2}$ the acceptor Na$^{+}$-ions
substitute for Ca$^{2+}$-ions and sit in the center of a square of
four Cu sites both above and below the outermost CuO$_2$ layer. The
attractive potential for the doped holes generated by a Na$^{+}$
acceptor does not break the local square symmetry.  The local square lattice symmetry broken
observed in the STM data is most pronounced at the in-plane oxygen sites
and for the lower voltage cutoff.  This has
raised an interesting question about the origin of the broken local square
symmetry in the STM experiments.

\begin{figure}[b]
\begin{center}
\includegraphics[width=7.4cm]{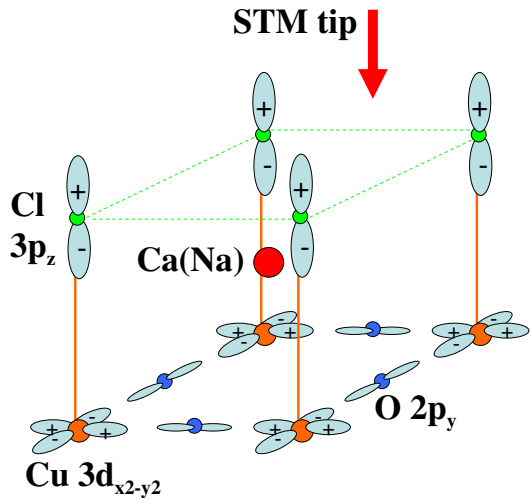}
\end{center}
\caption{Schematic crystal structure of
Ca$_{2-x}$Na$_{x}$CuO$_{2}$Cl$_{2}$ around a Ca$^{2+}$ ion or a
doped Na$^{+}$ ion.  The sign represents the phase of the Cu- and
Cl-orbital wavefunction.}
\end{figure}

In this letter we address the origin of this broken local symmetry.
We show that such broken symmetry states appear in the case of a
hole confined to a cluster of sites centered at a Na$^{+}$-acceptor.
Further this broken local symmetry shows up in the STM mapping
primarily as a modulation pattern above the O-sites in the outermost
CuO$_2$-layer.

We start by considering the STM tunneling process in detail in
Ca$_{2-x}$Na$_{x}$CuO$_{2}$Cl$_{2}$. As pointed out by Rice and
Tsunetsugu~\cite{Rice1}, the STM tip will couple primarily to the
3p$_z$-states of the outermost Cl layer but the matrix element for
an electron to tunnel directly into a hole located on the CuO$_2$
square directly underneath the Cl-ion vanishes by symmetry. The
holes move in a band of singlets in a state with
d$_{x^2-y^2}$-symmetry centered on a Cu-site but strongly hybridized
with the four nearest-neighboring (n.n.) O sites that surround each
Cu. The 3p$_z$-Cl states however hybridize with the hole states
centered on the four n.n. Cu sites. Each of these states hybridizes
either through the direct overlap of the 3p$_z$-Cl with
2p$_{x(y)}$-O orbitals and also through the 4s-Cu orbital on the
site directly below each Cl. Let $|\Psi_{0}^{1h} \rangle$ denote the
single hole ground state with energy $E^{1h}_0$, we can write the
amplitude to inject an electron in a 3p$_z$-Cl hole wavefunction as
a superposition in hole-states centered on the four n.n. Cu-site.
\begin{eqnarray}
p^{\dagger}_{Cl,\vec{i},{\sigma}} |\Psi_0^{1h} \rangle & \propto &
\sum_{\vec{\tau}}
 \langle \vec{i}, Cl| \vec{i}+ \vec{\tau}, Cu \rangle c^{\dagger}_{\vec{i}+\vec{\tau},\sigma}
|\Psi_{0}^{1h} \rangle \nonumber \\
& \propto & \sum_{\vec{\tau}} (-1)^{M_{\vec \tau}}c^{\dagger}_{\vec{i}+\vec{\tau},\sigma}
|\Psi_{0}^{1h} \rangle
\end{eqnarray}
where $p^{\dagger}_{Cl,\vec{i},{\sigma}}
(c^{\dagger}_{\vec{i}+\vec{\tau},\sigma})$ are creation operators
for electrons in the 3p$_z$-Cl orbital at site i (planar coordinate)
and the d-p hybridized orbital centered on the Cu-site at
$\vec{i}+\vec{\tau}$ in the outermost CuO$_2$ layer respectively,
$\vec{\tau}= \pm \hat x, \pm \hat y$ is a planar vector connecting
n.n. Cu sites. $\langle \vec{i}, Cl| \vec{i}+ \vec{\tau}, Cu
\rangle$ denotes the overlap between a 3p$_z$-Cl and the d-p
hybridized orbital centered at a n.n. Cu site. Assuming the square
lattice is four-fold rotational invariant, the overlaps are
independent of $\vec \tau$ except for their sign, which leads to the
final result in Eqn. (1). For $d_{x^2 -y^2}$ symmetry of the
Cu-orbital, we have $(-1)^M= -1$ for $\vec \tau = \pm \hat x$, and
$(-1)^M=+1$ for $\vec \tau =\pm \hat y$. Note that only the relative
phase is important.

Following the theory of STM tunneling processes developed by Tersoff
and Hamann~\cite{Tersoff}, we can write the differential conductance
at voltage V at $\vec r$, the center of curvature of the tip,
\begin{equation}
\frac{dI(\vec{r})}{dV} \propto \sum_{\sigma,m} |\langle
m|a^{\dagger}_{\vec{r},\sigma}|\Psi_{0}^{1h} \rangle|^2 \delta_{E_m - E^{1h}_0, \omega}
\end{equation}
where $a^{\dagger}_{\vec{r},\sigma}$ is the electron creation
operator at position $\vec{r}$, $|m \rangle$ are eigenstates of the
half-filled system with energy $E_m$, and $\omega =eV$. When the tip
is scanned from above the Cl site at $\vec{i}$ to a neighboring
site,  $\vec{i}+\vec{\tau'}$, (planar coordinates), we have
\begin{equation}
a_{\vec{r},\sigma}^{\dagger} |\Psi_{0}^{1h} \rangle= [ \langle \vec{r}|\vec{i}, Cl \rangle
p_{Cl, \vec{i}, \sigma}^{\dagger} + \langle
\vec{r}|\vec{i}+\vec{\tau'}, Cl \rangle p_{Cl, \vec{i}+\vec{\tau'},
\sigma}^{\dagger}]|\Psi_{0}^{1h} \rangle.
\end{equation}
The integrated current at $\vec r$ up to a positive
voltage $V$ is then,
\begin{equation}
I(\vec r, \omega) = A \sum_{\sigma,m}|\langle
m| \sum_{\vec \tau} (-1)^{M_{\vec \tau}} [\langle \vec r | \vec i, Cl \rangle c^{\dagger}_{\vec i +\vec \tau,\sigma}
+ \langle \vec r | \vec i +\vec \tau', Cl \rangle c^{\dagger}_{\vec i + \vec \tau'+\vec \tau,\sigma}
|\Psi_{0}^{1h} \rangle|^2 \Theta(\omega -E_m +E^{1h}_0)
\end{equation}
where $A$ is a constant, and $\Theta$ is a step function.
The variation of the tunneling current as one scans between two n.n.
Cl sites will therefore be sensitive to the relative phase to inject
electrons on n.n. Cu sites in $|\Psi_{0}^{1h} \rangle$. To this end
we examine the lowest energy single-hole states for clusters up to
16 sites including an attractive potential for the center square.

The two-dimensional single-band $t$-$t'$-$J$ model with the on-site
impurity potential on four Cu sites of the center square around the
Na$^{+}$ dopant is defined by the Hamiltonian,
\begin{eqnarray}
{\cal H} &=& - \sum_{i,\sigma} \epsilon _{i} c^{\dag}_{i\sigma}
c_{i\sigma} -t\sum_{\langle i,j \rangle \sigma} (c^{\dag}_{i\sigma} c_{j\sigma} + {\rm h.c.})
\nonumber \\
& & -t'\sum_{ \langle \langle i,j \rangle \rangle \sigma}
(c^{\dag}_{i\sigma} c_{j\sigma} + {\rm h.c.}) +
J\sum_{\langle i,j\rangle} \mbox{\boldmath $S_{i}\cdot S_{j}$}.
 \label{hamil}
\end{eqnarray}
A constraint of no double electron occupation is implied.  $\langle
i,j\rangle$ and $\langle \langle i,j \rangle \rangle$ refer to n.n.
and next n.n. sites $i$ and $j$. The onsite potential of the four
n.n. Cu sites around the sodium dopant is $\epsilon
_{i}=\epsilon_{s}$ while $\epsilon _{i}=0$ for the rest of sites.
The single hole impurity state of the Hamiltonian (4) at $t'=0$ was
studied previously by using exact diagonalization method for small
clusters by Von Szczepanski {\em et al.}~\cite{tjimp1}, Rabe and
Bhatt ~\cite{tjimp2}, and Gooding~\cite{tjimp3}. The ground state
was found to be orbitally doubly degenerate ($S=1/2$) in certain
parameter regime ~\cite{tjimp2,tjimp3,footnote}. When the next n.n.
hopping integral $t'$ is included, our results below show that the
doubly degenerate S=1/2 states is lowest in energy within a
reasonable parameter range. The symmetry of the lowest energy state
however is sensitive to boundary condition and parameters. In our
calculations, we choose $t=1$ and $J=0.3$, suitable for the
cuprates, and study the ground state as a function of $t'$ and
$\epsilon_s$. We consider the clusters of 12- and 16-sites as
illustrated in Fig. 2.

\begin{figure}[t]
\begin{center}
\includegraphics[width=12cm]{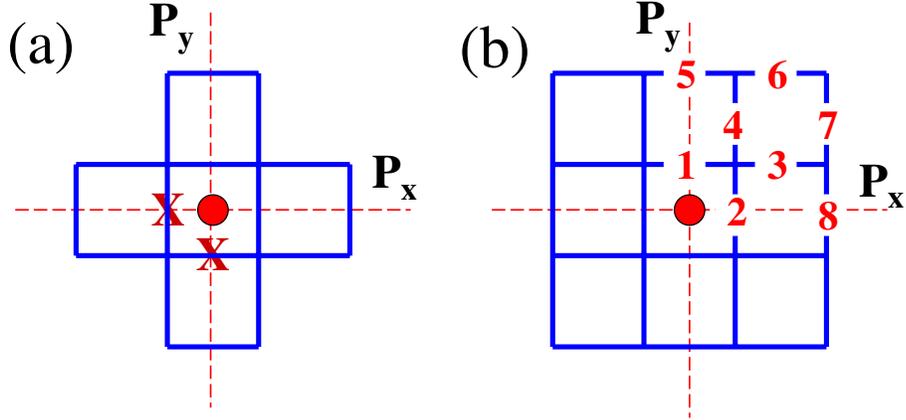}
\end{center}
\caption{The configuration of 12-site (a) and 16-site cluster (b).
The dashed line corresponds to the reflection axis for $P_{x(y)}$.
The red cross marks the tip position on the middle point of two n.n.
Cu sites along the x- and y-axes. In (b), eight independent bonds
are labeled with numbers.}
\end{figure}

Exact diagonalization calculations of the $t$-$t'$-$J$ model on
small clusters have been
reported~\cite{Dagotto1,Leung1,Hass,Troyer,TXiang,TKLee}. For a
16-(4$\times$4) site cluster with periodic boundary condition (PBC),
the ground state of a single hole has a four-fold symmetry with hole
momentum $(\pm \pi/2,\pm \pi/2)$ if $t' <0$, and a two-fold symmetry
with hole momentum $(\pi,0)$ or $(0,\pi)$ if $t'>0$.

The presence of an impurity potential on the lightly doped $t$-$J$
model affects not only the local charge and spin distributions but
also the symmetry of the ground state
wavefunction~\cite{tjimp1,tjimp2,tjimp3}. In this study, we focus on
the reflection symmetries of a two-dimensional square lattice with
respect to $x$- and $y$-axes passing through the sodium dopant
($P_x$ and $P_y$ respectively) and on the parity $P_xP_y$. Since
$[P_{x(y)},H]=0$, we may classify states according to the quantum
numbers of $P_x$, $P_y$. We denote the state with $(P_x=+1,P_y=+1)$
as phase (I), and doubly degenerate state $(+1,-1)$,\,and  $(-1,+1)$
as phase (II), and $(-1,-1)$ as phase (III). In Fig. 3(a) and 3(b),
we show the ground state phase diagram in terms of its reflection
symmetry obtained by exact diagonalization. As we can see, there is
a regime with the doubly degenerate ground state. Previous band
structure calculation~\cite{LDA} and angle resolved photoemission
experiments~\cite{ARPES} both indicate a rather weak $t' \approx
-0.1$ for Ca$_{2-x}$Na$_{x}$CuO$_{2}$Cl$_{2}$. Thus we expect the
existence of doubly degenerate ground state in this doped material.
We note that the four low-lying states with distinct reflection
symmetries are very close in energy. Small perturbation of lattice
geometry, boundary condition, $t'$ or $\epsilon_s$ may change
symmetry of the ground state. In Fig. 3(c), we show the energy level
evolution as $\epsilon_s$ changes for a fixed value of $t'=-0.12$.

\begin{figure}[tb]
\begin{center}
\includegraphics[width=15cm]{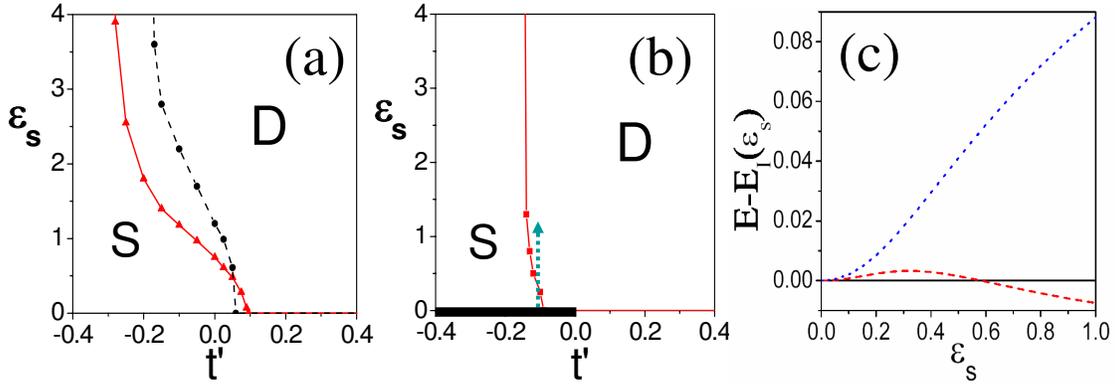}
\end{center}
\caption{Ground state phase diagram in the parameter space of the
next n.n. hopping integral $t'$ and the sodium impurity potential
$\epsilon_s$ in (a) 12-site cluster (black solid line) and 16-site
cluster (red dashed line) with open boundary condition, (b) 16-site
cluster with PBC. Regime D: the doubly degenerate state; regime S:
the singly degenerate state. The thick black line in (b) corresponds
to the four-fold symmetric states at $\epsilon_s=0$. (c) Energy
levels as functions of $\epsilon_s$ at $t'=-0.12$ as indicated by
the dashed green line in (b). $E_I$ denotes the energy in phase I.
The dashed line is for doubly degenerate states while the dotted
line is for phase III.}
\end{figure}

We next examine the effect of the rotational symmetry broken ground
state to the STM measurements. We consider the system to be in one
of the two degenerate states, say in the state of $P_x =1, P_y=-1$.
The broken symmetry may be caused by quadrupole interaction of two
single-hole states or by other couplings. We remark that the local
density of states at the Cu-sites remains rotational symmetric in
this state. However, because of the interference of different
Cu-sites to the integrated differential conductance in Eqn. (4), the
STM tip at a position between two Cu-sites will break the rotational
symmetry. To be specific, we consider the tip position above the two
geometrically  symmetric points at halfway between the Cu-site
$\vec{i}$ and its neighboring sites $\vec{i}+ \hat x$ or $\vec{i}+
\hat y$. The sodium impurity is located at $\vec{i} + \hat x/2 +\hat
y/2$. Note that $\langle \vec r|\vec i, Cl \rangle = \langle \vec
r|\vec i +\tau', Cl \rangle $ for the mid point $\vec r$, we obtain
the expression for the integrated differential conductance $I^{x
(y)}(\omega) = I(\frac{\hat x}{2} (\frac{\hat y}{2}), \omega)$ at
position $\vec i + \frac{\hat x}{2} (\frac{\hat y}{2})$ with the
cut-off energy $\omega$,
\begin{eqnarray}
I^{x (y)}(\omega) &=& \sum_{\sigma} I^{x (y)}_{\sigma}(\omega), \nonumber \\
I^{x (y)}_{\sigma}(\omega) &=& A_0 \sum_{m} |\langle
m| \sum_{\vec{\tau}} (-1)^{M_{\vec{\tau}}}
(c^{\dagger}_{\vec{i}+\vec{\tau},\sigma} +
c^{\dagger}_{\vec{i}+\hat x (\hat y) +\vec{\tau},\sigma})|\Psi_0^{1h}
\rangle|^2 \Theta(\omega - E_m+E^{1h}_0),
\end{eqnarray}
where $I_{\sigma}$ is the spin-dependent integrated differential
conductance, and the Cu-site index $\vec i$ has been dropped for
simplicity. The conductance can be calculated in a small cluster. We
diagonalize the hamiltonian $H$ in Eqn. (5). exactly and find the
ground state of the single hole and also all the eigenstates $|m
\rangle$ and the corresponding energies $E_m$ at the half-filling,
which allows us to calculate all the matrix elements in Eqn. (6).
Note that $I^x(\omega) = I^y(\omega)$ by symmetry if $|\Psi_0^{1h}
\rangle$ is non-degenerate. In the parameter regime where
$|\Psi_0^{1h} \rangle$ is doubly degenerate (see Fig. 2),  $I^x$ and
$I^y$ can be different. In Fig. 4(a), we plot $I^x(\omega)$ and
$I^y(\omega)$ as functions of the cutoff energy $\omega$ obtained
from the 12-site cluster calculations, where the hole groundstate
has symmetry  $P_x=+1, P_y =-1$ and has a spin S$_z$=1/2. The
asymmetry of $I$ between the two tip scans is apparent and is more
pronounced at a lower energy cutoff and becomes weaker at a high
energy cutoff. To see the asymmetry more clearly, we also plot the
ratio of the conductances at the two tip points in the same figure,
\begin{equation}
\eta(\omega) = I^x(\omega)/I^y(\omega).
\end{equation}

\begin{figure}[t]
\begin{center}
\includegraphics[width=9.5cm]{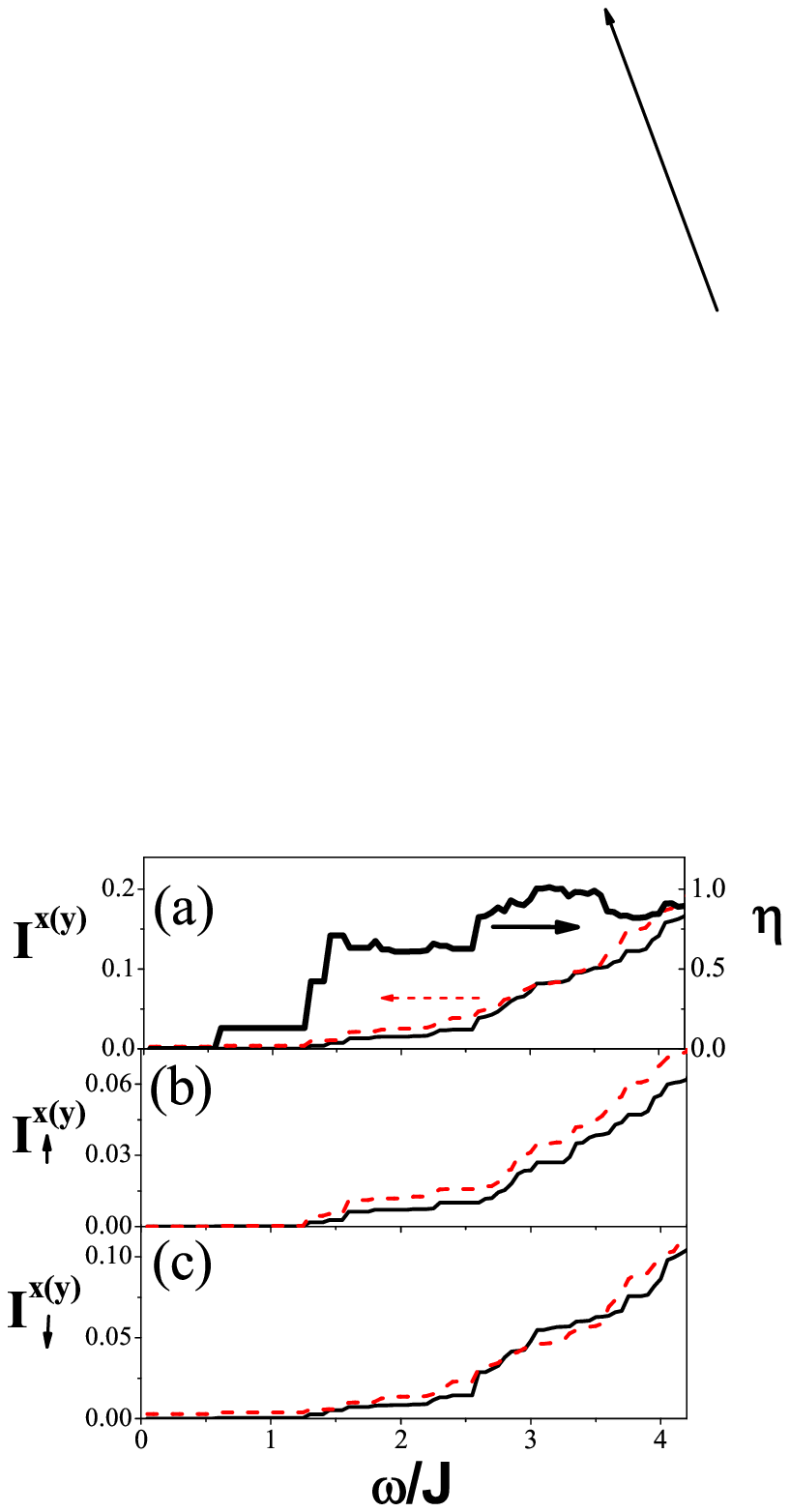}
\end{center}
\caption{The total integrated differential conductance on two cross
(X) sites in Fig. 2a along the $x$- (solid line) and along the
$y$-axis (dashed line) as functions of the cut-off energy $\omega$
together with the anisotropy ratio $\eta$ (thick solid line). The
results for spin-up channel (b),  and spin-down channel (c). In the
calculations (b), and (c), the single hole state has a spin
S$_z$=1/2. }
\end{figure}

It is interesting to note the strong cutoff energy dependence of the
asymmetry in the conductance. For example, $\eta \approx 0.3$ for
$\omega \approx J$, and $\eta \approx 0.85$ for $\omega =4 J$. At
$\omega \rightarrow \infty $, $\eta \approx 1.04$, which shows a
very weak asymmetry. Our theoretical results are consistent with the
recent STM observation~\cite{STM2}, where the strong asymmetry has
been found at a lower energy 150 meV ($\sim J$), and the asymmetry
becomes much weaker or unobservable at a high energy 600 meV ($\sim
4J$). In Fig. 4(b) and 4(c), we show the spin-dependent conductances
$I_{\sigma}(\tau'/2, \omega)$ as functions of $\omega$ in different
spin channels.  Our theory predicts that the asymmetry is
spin-dependent. At high energy cutoff limit, we find the ratio
$I^x_{\uparrow}/I^y_{\uparrow} \sim 1.23$ in the spin-up channel and
$I^x_{\downarrow}/I^y_{\downarrow} \sim 0.88$ in the spin-down
channel.  The asymmetry compensates each other in the two different
spin channels, which gives a weaker asymmetry in the total ratio
$I^x/I^y \sim 1.04$.  We remark that rapid spin flip processes
around the sodium impurity may make the spin-resolved STM
experiments more difficult to observe. However, at very underdoped
samples, the spin may be frozen around the sodium impurity, which
should provide the possibility for the spin-resolved STM
experiments.

To better understand our results, we analyze the groundstate and the
conductance in a four-site cluster system. The 4-site cluster of the
Hubbard model with a single hole was previously studied by Altman
and Auerbach~\cite{Auerbach}. They have found that for not very
large on-site Coulomb repulsion $U$, the ground state is two-fold
degenerate with spin-1/2.  We here study the single hole ground
state of the Hamiltonian Eqn. (5), and focus on the spin-1/2 sector,
which is relevant for not too small ratio of $J/t$. Note that the
impurity potential does not play any role in the 4-site cluster. The
ground state is doubly degenerate with the energy given by
$E_g=-J/2-\sqrt{3t^2+(t'+J/2)^2}$, with $tan
\gamma=\sqrt{3}t/(J+t'+E_g)$.  The two degenerate states can be
classified by symmetry $P_x=+1, P_y=-1$ and $P_x=- 1, P_y=+1$.
Assuming the former to be ground state, we find that apart from an
overall prefactor, $I^x_{\uparrow} (\omega) = \frac{4}{3}
\cos^{2}\gamma \Theta (\omega-J)$, $I^x_{\downarrow} (\omega) =
\frac{2}{3} \cos^{2}\gamma \Theta(\omega-J)$, $I^y_{\uparrow}
(\omega)=0$ and $I^y_{\downarrow} (\omega) = \cos^{2}\gamma
\Theta(\omega) + \sin^{2}\gamma \Theta(\omega-2J)$. Due to the
nature of the many body wavefunction, the conductance in Eqn.(6) may
involve either destructive or constructive interference among
different components. This effect is most drastic in
$I^y_{\uparrow}$ with an exact cancelation between them. Summing
over spins, we have, $I^x(\omega)=I^x_{\uparrow}+ I^x_{\downarrow} =
2 \cos^2\gamma \Theta(\omega-J)$ while $I^y (\omega) = \cos^2\gamma
\Theta(\omega) +  \sin^2\gamma \Theta(\omega-2J)$. It is obvious
that $\eta(\omega)$ becomes infinite for $\omega \in (0,J)$,
$\eta(\omega)=1/2$ for $\omega \in (J,2J)$ while $\eta(\omega)=2
\cos^{2}\gamma$ for $\omega
> 2J$. At the limit of the large bias voltage, we
have $\eta = 2 \cos^{2}\gamma$. For $t'/t=-0.1$,
 $\eta \sim 0.97$. Therefore, the asymmetry becomes rather weak
with the increase of bias voltage. The ratio of the integrated
conductances in the ground state with symmetry $P_x=-1, P_y=+1$ can
be obtained by the interchange symmetry of $x$ and $y$, and is given
by $1/\eta$. We have also calculated the integrated conductance for
the hole injection tunneling, and found also rotational asymmetric.
The ratio of the conductances between the electron injected and hole
injected is also asymmetric.

\begin{table}
\begin{tabular}{|c|c|c|c|c|c|c|c|c|}\hline

 bond index & $1 $  & $ 2$ & $3$ & $4$  & $5$ & $6$ & $7$ & $8$\\
   \cline{1-9}
 $\langle C_i^{\dagger}C_j \rangle$  &0.342  &0.328  & 0.122   & 0.129 & 0.020  &
 0.014 & 0.012 & 0.016\\
  \cline{1-9}
 $-\langle S_i S_j \rangle$  & 0.167  &0.074  & 0.161   & 0.167 &
 0.329  & 0.336 & 0.347 & 0.334  \\
  \hline
\end{tabular}
\caption{
The hopping integral and spin-spin correlation function for various
bonds in a 16-site cluster with PBC for $t'=-0.1$ and
$\epsilon_s=1.0$. The bond indices are labeled in Fig. 2(b).}
\end{table}

In the rotational symmetry broken state we considered above, the
local charge density at each of four n.n. Cu sites around sodium
dopant is uniform, but the n.n. Cu-Cu bondings are not equal. For
instance, in the ground state with $P_x=+1, P_y =-1$ of the 16-site
cluster, we show the expectation value of hopping integral $\langle
C_i^{\dagger}C_j \rangle$ as well as spin-spin correlation function
$\langle S_i S_j \rangle$ for various bonds, depicted in Table I.
The hopping integral along bond 1 is slightly stronger than that of
bond 2 while the spin-spin correlation function exhibits a much
stronger asymmetry. The impurity may lead to localization of holes
around this impurity which may result in a larger hopping integral
and weaker spin-spin correlation function around this impurity.
Close to the lattice boundary, since the hole density becomes very
small for large impurity scattering strength, the spin-spin
correlation function may approach to -0.34 and hopping integral
vanishes. Our results are consistent with previous calculations. The
strong variations of hopping integral and spin-spin correlation
around the impurity may lead to the formation of local lattice
distortion and the appearance of spin-lattice polaron due to the
electron-phonon coupling~\cite{polaron}, which may induce further
rotational asymmetry. We speculate that this could be one of the
reasons for the inhomogeneity in underdoped
Ca$_{2-x}$Na$_{x}$CuO$_{2}$Cl$_{2}$.

The authors thank Y. Kohsaka and S. Davis for discussions of their
experiments. YC especially thanks P.W. Leung, T.K. Lee, H.
Tsunetsugu, and M. Troyer for numerical assistance. YC and FCZ are
supported by RGC grants from the Hong Kong SAR government, and TMR
by the MANEP program of the Swiss Nationalfonds.


\begin{thebibliography}{}
\bibitem{footnote} The double degenerate ground state was not found in Ref.~\cite{tjimp1},
due to the lower symmetry of the cluster they considered.
\end{thebibliography}
\end{document}